\documentclass[12pt,letterpaper]{JHEP3}
\usepackage{graphics}
\usepackage{epsfig}

\title{Covariant entropy conjecture and concordance cosmological models}
\author{Song He\\
School of Physics, Peking University, Beijing, 100871, China\\
\email{hesong@pku.edu.cn}}
\author{Hongbao Zhang\\
Perimeter
Institute for Theoretical Physics, Waterloo, Ontario, N2L 2Y5,
Canada
\\Department of Astronomy, Beijing Normal University, Beijing,
100875, China\\
\email{hzhang@perimeterinstitute.ca}}

\abstract{Recently a covariant entropy conjecture has been proposed for
dynamical horizons. We apply this conjecture to concordance
cosmological models, namely, those cosmological models filled with
perfect fluids, in the presence of a positive cosmological constant.
As a result, we find this conjecture has a severe constraint power.
Not only does this conjecture rule out those cosmological models
disfavored by the anthropic principle, but also it imposes an upper
bound $10^{-60}$ on the cosmological constant for our own universe,
which thus provides an alternative macroscopic perspective for
understanding the long-standing cosmological constant problem.}
\begin{document}
\section{Introduction}
The notion of dynamical horizons, free of the global and
teleological deficit of event horizons, was developed quasi-locally
and its properties were extensively investigated, where, in
particular, the first and second laws were generalized to the black
hole dynamical horizon\cite{AK0,AK1,AK2}. Following this line further,
we have recently proposed a covariant entropy bound formulation of
an analogous generalized second law of thermodynamics on the black
hole dynamical horizon and its generalization to cosmological
dynamical horizons in FRW universes has also been
conjectured\cite{HZ1,HZ2}. Moreover, with the assumption that the
dominant energy condition and adiabatical evolution hold for matter,
its validity has been confirmed in both Vaidya black holes and FRW
universes full of matter with a fixed state of equation, no matter
whether the spatial geometry is open, flat, or closed\cite{HZ1,HZ2}.
All of these results suggest that dynamical horizons may also have
an interpretation of thermodynamics. However, even though it turns
out to be not true, as inferred above, our proposal itself can still
be viewed as a covariant entropy bound conjecture on dynamical
horizons and there may be some deep reasons for its validity. In
fact, the conjecture is motivated partly by Bousso entropy bound
conjecture\cite{Bousso0,Bousso1,Bousso2,Bousso3}, and its
strengthened form suggested by Flanagan, Marolf, and
Wald\cite{Flanagan}. In particular, when the dynamical horizon is spacelike, our conjecture has been proved on the basis of the
strengthened form\cite{GW}.
These various entropy bound conjectures,
including ours, can also be interpreted as a statement of the so
called holographic principle, which is believed to be manifest in an
underlying quantum gravity\cite{Hooft,Susskind}.

Due to its success in many respects and justification as a possible
fundamental principle, it is intriguing to know if our conjecture is
able to rule out some types of cosmological models or to provide
somewhat particular constraints on certain significant
cosmological parameters of acceptable models, such as the
cosmological constant in our own universe, which is just what we
shall address in this paper. As a first step in this direction, we
shall here apply our conjecture to those cosmological models with a
positive cosmological constant plus the matter content satisfying
the dominant energy condition, which our own universe belongs to.
Remarkably, it is found that our conjecture rules out all of those
cosmological models with $0<w\leq 1$ for the matter content. In
addition, although our own universe satisfies our conjecture as it
should, with a safety margin of 30 orders of magnitude,
%not only does
our conjecture alleviates the cosmological constant problem by
imposing an upper bound on the cosmological constant.
% but also it
%indicates there may be a certain profound connection among the
%cosmological constant and origin of mass.

Planck units are used here, i.e., $c=G=\hbar=k=1$, where $c$ is the
speed of light, $G$ is the Newton constant, $\hbar$ is the Planck constant,
and $k$ is the Boltzmann constant. Notation and conventions follow
\cite{Wald}
\section{Covariant entropy conjecture associated with cosmological dynamical horizon}
In terms of the conformal time and comoving coordinate, the FRW
metric takes the form
\begin{equation}
ds^2=a^2(\eta)[-d\eta^2+d\chi^2+f^2(\chi)(d\theta^2+\sin^2\theta
d\phi^2)],
\end{equation}
which describes homogeneous and isotropic universes, including, to a
good degree of approximation, the portion we have seen of our own
universe. Here $f(\chi)=\sinh\chi,\chi,\sin\chi$ correspond to open,
flat, and closed universes, respectively.

With the line element above, the future directed null congruences
orthogonal to an arbitrary sphere characterized by some value of
$(\eta,\chi)$ can be chosen as
\begin{equation}
k^a_{\pm}=\frac{1}{2}[(\frac{\partial}{\partial\eta})^a\pm(\frac{\partial}{\partial\chi})^a],
\end{equation}
whose expansions are given by\cite{Ben}
\begin{equation}
\theta_{\pm}=\frac{\dot{a}}{a}\pm\frac{f'}{f},\label{expansion}
\end{equation}
where the dot(prime) denotes differentiation with respect to
$\eta(\chi)$, and the sign $+(-)$ represents the null congruence
directed at larger(smaller) values of $\chi$. Note that the second
term is given by $\coth\chi,\frac{1}{\chi},\cot\chi$ for open, flat,
and closed universes, respectively. In particular, this term diverges
when $\chi\rightarrow 0$, and it also diverges when $\chi\rightarrow
\pi$ for a closed universe.

Then the cosmological dynamical horizon can be defined geometrically as
a three-dimensional hypersurface foliated by those spheres at which
at least there exists one orthogonal null congruence with vanishing
expansion. Thus the cosmological dynamical horizon $\chi_c(\eta)$ is
identified by solving the equation
\begin{equation}
h=\pm\frac{f'}{f}|_{\chi=\chi_c},\label{horizon}
\end{equation}
where $h\equiv\frac{\dot{a}}{a}$. There is one solution for open and
flat universes, while for a closed universe, there are generally two
solutions, which are symmetrically related to each other by
$\chi_c^2(\eta)=\pi-\chi_c^1(\eta)$. Now the covariant entropy
conjecture on cosmological dynamical horizon can be addressed as
follows\cite{HZ1}: {\it{Let $A(\eta)$ be the area of the cosmological
dynamical horizon at the conformal time $\eta$, then the entropy
flux $S$ through the cosmological dynamical horizon between the
conformal times $\eta$ and $\eta'$ must satisfy
$S\leq\frac{|A(\eta)-A(\eta')|}{4}$ if the dominant energy condition
holds for matter}}.

It is noteworthy that unlike for the black hole dynamical
horizon, we have no restriction on the signature of cosmological
dynamical horizons in our definition, which means our cosmological
dynamical horizons can be spacelike, null, or timelike\cite{Ben}.
This extension, however, does not ruin the validity of our covariant
entropy conjecture in the cosmological context. As demonstrated in
\cite{HZ1}, our conjecture always holds regardless of the
signature of cosmological dynamical horizons.

\section{Acceptable cosmological models constrained by covariant entropy conjecture}
If as usual the matter content of FRW universes is assumed to be
described by the perfect fluid with energy momentum tensor
\begin{eqnarray}
T_{ab}&=&a^2(\eta)\{\rho(\eta) (d\eta)_a(d\eta)_b+p(\eta)[(d\chi)_a
d(\chi)_b\nonumber\\
&&+f^2(\chi)((d\theta)_a(d\theta)_b+\sin^2\theta(d\phi)_a(d\phi)_b)]\},
\end{eqnarray}
then by the Einstein equation with a positive cosmological constant
$\Lambda$, we have
\begin{eqnarray}
3(h^2+K)&=&8\pi\rho a^2+\Lambda a^2,\label{time}\\
-(h^2+2\dot{h}+K)&=&8\pi p a^2-\Lambda a^2,\label{space}
\end{eqnarray}
where $K=1,0,-1$ correspond to open, flat, and closed universes,
respectively. From here, we can further obtain
\begin{eqnarray}
\dot{h}&=&-\frac{4\pi}{3}[(1+3w)\rho-\lambda] a^2,\label{derivative}\\
h^2+K-\dot{h}&=&4\pi (1+w)\rho a^2,\label{combination}
\end{eqnarray}
and the energy momentum conservation equation
\begin{equation}
\dot{\rho}=-3(1+w)\rho h,\label{conservation}
\end{equation}
where $-1\leq w\equiv\frac{p}{\rho}\leq1$ due to the dominant energy
condition, and $\lambda\equiv\frac{\Lambda}{4\pi}$.

To proceed, we further assume that the evolution of FRW universes is
adiabatical, which implies the conservation of the entropy current
associated with the matter, i.e., $\nabla_as^a$=0. Hence the
entropy current can be formulated as
\begin{equation}
s^a=\frac{s}{a^4}(\frac{\partial}{\partial\eta})^a,
\end{equation}
where $s$ is actually the ordinary comoving entropy density,
constant in space and time.

\begin{figure}[htb!]
\begin{center}
\includegraphics[clip=truth,width=1.2\columnwidth]{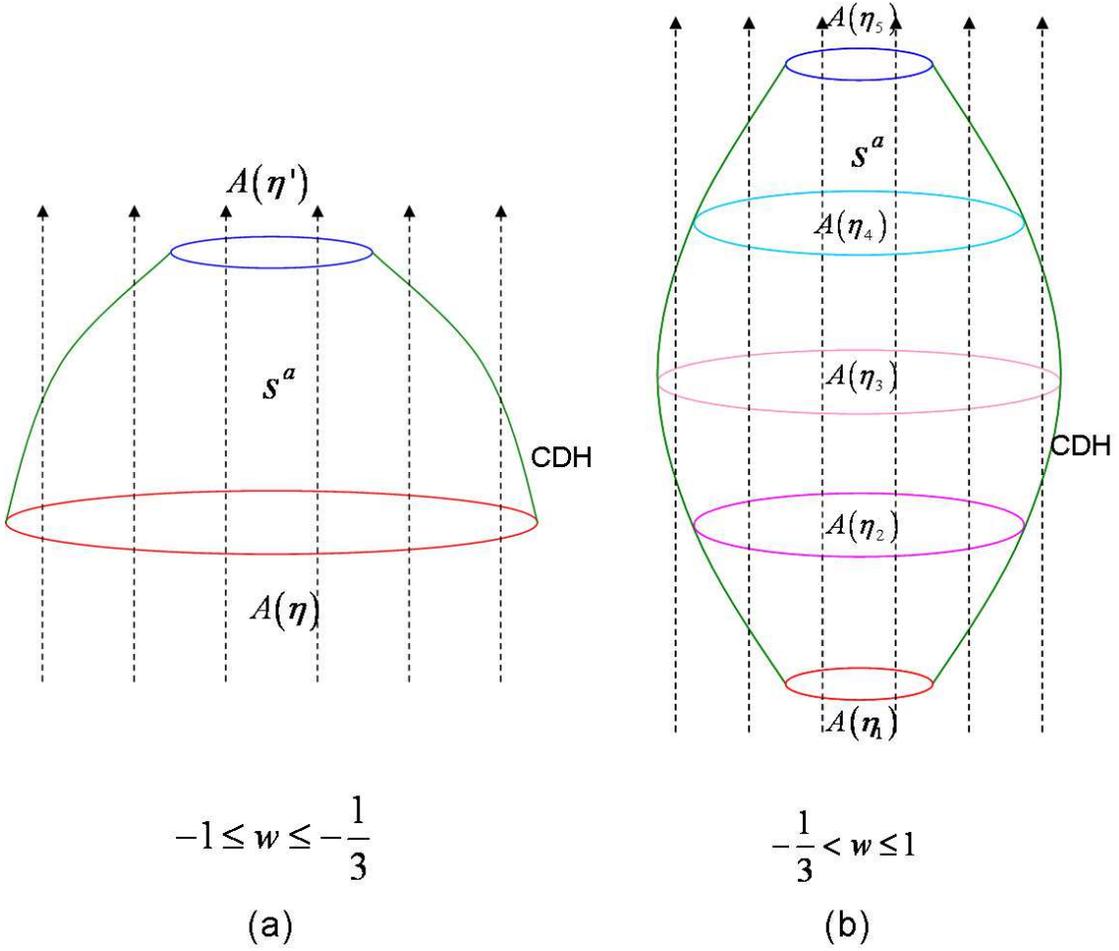}
\end{center}
\caption {The cosmological dynamical horizon with the entropy current
flowing through it in the conformal coordinate. When $\dot{\chi}_c\leq
0$, the entropy current flows across the cosmological dynamical horizon
from the interior region to the exterior one, while it flows from the
exterior region to the interior one for $\dot{\chi}_c\geq 0$.}
\label{cdh}
\end{figure}

On the other hand, according to eq.(\ref{horizon}), we have
\begin{equation}
\dot{\chi_c}=\mp\frac{\dot{h}}{h^2+K}.
\end{equation}
Note that at any moment the area of the cosmological dynamical horizon
is give by $A=4\pi a^2 f^2(\chi_c)$. Accordingly we can work out its
time derivative, i.e.,
\begin{eqnarray}
\dot{A}&=&8\pi
a^2f^2(\chi_c)(h+\frac{\dot{f}(\chi_c)}{f(\chi_c)})=8\pi
a^2f^2(\chi_c)(h+\frac{f'(\chi_c)\dot{\chi}_c}{f(\chi_c)})\nonumber\\
&=&8\pi a^2f^2(\chi_c)\frac{h}{h^2+K}(h^2+K-\dot{h}).
\end{eqnarray}
Obviously, by eq.(\ref{time}) and eq.(\ref{combination}), the
increase or decrease of area with time only depends on whether the
universe is expanding or contracting. In what follows we shall only
focus on the expanding universes, i.e., $h\geq 0$, where the
corresponding area monotonically increases with the evolution of
time.

We shall now check whether our conjecture is satisfied for those
universes mentioned above. However, as demonstrated in figure
\ref{cdh}, it is noteworthy that there is an obvious difference
between $\dot{\chi}_c\leq 0$ and $\dot{\chi}_c\geq 0$. Thus
employing the conservation of the entropy current and Gauss theorem,
our conjecture can be equivalently expressed as
\begin{equation}
\frac{\dot{A}}{4}+\dot{S}\geq 0
\end{equation}
for $\dot{\chi}_c\leq 0$($\dot{h}\geq 0$), and
\begin{equation}
\frac{\dot{A}}{4}-\dot{S}\geq 0
\end{equation}
for $\dot{\chi}_c\geq 0$($\dot{h}\leq 0$)\footnote{As explained in
\cite{HZ1}, this can be reasonably taken as the generalized
second law of thermodynamics in the cosmological context.}. Here $S$
denotes the entropy flux through the interior region $\chi\leq
\chi_c$, given by
\begin{equation}
S=4\pi s\int_0^{\chi_c}d\chi f^2(\chi),
\end{equation}
whereby we have
\begin{equation}
\dot{S}=4\pi sf^2(\chi_c)\dot{\chi}_c=-4\pi
sf^2(\chi_c)\frac{\dot{h}}{h^2+K}.
\end{equation}
So for $-1\leq w\leq-\frac{1}{3}$, our conjecture requires
\begin{equation}
s\leq \frac{\sqrt{3(8\pi\rho a^2+\Lambda a^2-3K)}(1+w)\rho
a^2}{2[\lambda-(1+3w)\rho]},\label{inequality}
\end{equation}
where eq.(\ref{time}), eq.(\ref{derivative}), and
eq.(\ref{combination}) have been used. Furthermore, by
eq.(\ref{conservation}), we know the energy density $\rho\propto
a^{-3(1+w)}$. Thus the rhs of the above inequality is an
increasing function of $a$, which is similar to our previous result
without the cosmological constant\cite{HZ1}. That is to say, the
bound will hold forever once it holds at some early time, where
classical general relativity becomes an effective theory for 
successfully describing the evolution of universes. This is expected
since the cosmological constant corresponds to $w=-1$ which falls
into the branch $-1\leq w\leq-\frac{1}{3}$ and a combination of the
cosmological constant with the matter content in this branch will
not change the situation to violate our conjecture.

Similarly, for the case of $-\frac{1}{3}<w\leq 1$, our conjecture is
equivalent to
\begin{equation}
s\leq \frac{\sqrt{3(8\pi\rho a^2+\Lambda a^2-3K)}(1+w)\rho
a^2}{2[(1+3w)\rho-\lambda]}\label{early}
\end{equation}
when $(1+3w)\rho>\lambda$, and
\begin{equation}
s\leq \frac{\sqrt{3(8\pi\rho a^2+\Lambda a^2-3K)}(1+w)\rho
a^2}{2[\lambda-(1+3w)\rho]}\label{late}
\end{equation}
when $(1+3w)\rho<\lambda$, which apparently corresponds to later
stages of expanding universes. For simplicity but without loss of
generalization, we shall restrain our later discussions to the flat
case, i.e., $K=0$. Then after a straightforward calculation, we find
that the rhs of inequality (\ref{early}) is an increasing
function of $a$. However, the rhs of inequality (\ref{late}) is a
decreasing function of $a$ for $0\leq w\leq 1$; on the other hand,
for $-\frac{1}{3}<w<0$, it decreases with $a$ in the region of
$\lambda<\frac{1+\frac{2}{\sqrt{1+3w}}}{1-\sqrt{1+3w}}(1+3w)\rho$,
and then increases with $a$ in the region of
$\lambda>\frac{1+\frac{2}{\sqrt{1+3w}}}{1-\sqrt{1+3w}}(1+3w)\rho$.
Obviously, here we see a very different situation from our previous
one. Namely, for $-\frac{1}{3}<w\leq 1$, the bound may not be
guaranteed to hold at some late stages even if it holds at
reasonable early stages. Speaking specifically, for
$-\frac{1}{3}<w\leq 0$, whether the bound will hold depends
critically on whether the inequality (\ref{late}) will hold at the
point where the minimum for the rhs occurs. What is worse, for
$0<w\leq 1$, the bound is doomed to be violated at late stages of
the evolution of expanding universes since the rhs approaches zero
as $a\rightarrow\infty$ by $\rho\propto a^{-3(1+w)}$. However, one
may regard this observation as an indication that some types of
cosmological models can be ruled out by our conjecture. Putting this
another way, if we take this viewpoint seriously, we can claim that
our conjecture rules out those cosmological models with a mixed
content of a positive cosmological constant and matter satisfying a
fixed equation of state of $0<w\leq 1$.
%Especially, Penrose's
%recently proposed conformal cyclic cosmology is practically excluded
%by our conjecture since it is assumed that the massless particles
%will be the final matter content in the very remote future
%\cite{Penrose}.

Although all the discussion above is confined to only one kind of
matter content besides a positive cosmological constant, it is easy
to generalize to a more realistic cosmological model where
various types of matter are combined with a positive cosmological
constant. In particular, for the currently favored concordance
$\Lambda$CDM model, where
$\rho=\frac{\rho_r^0a_0^4}{a^{4}}+\frac{\rho_m^0a_0^3}{a^{3}}$,
on the basis of eq.(\ref{late}) and taking into account that the dust has
been the dominant matter content, our conjecture requires
\begin{equation}
s\sqrt{\Lambda}\leq 2\sqrt{3}\pi\rho^0_m,
\end{equation}
where $\rho^0_m$ denotes the energy density of dust today, and the
present scale factor is set to be $a_0=1$, which implies that $s$
represents the entropy density today. This is a remarkable result
because it establishes a novel relation governing the cosmological
constant, matter entropy density and dust energy density. For our
own universe, as is well known, $\Lambda\sim 10^{-120}$ and
$\rho^0_m\sim\frac{1}{3}\Lambda$, so from our conjecture it follows that the
present entropy density should be less than $10^{-60}$, which is
satisfied with a wide safety margin, since the realistic entropy
density is around of order $10^{-90}$ today. That is to say, our
conjecture supports the existence of our own universe as it should
do. On the other hand, if we take the present matter entropy density
and dust energy density as input data, our conjecture gives an upper
bound on the cosmological constant, i.e., $\Lambda<10^{-60}$, which
obviously alleviates the cosmological constant problem of the
cosmological constant being so small in Planck units. Last but not
least, the presence of cosmological constant, albeit small, appears
to be in favor of a bio-friendly universe: to have our conjecture
satisfied, there should be dust matter in our universe, which is
assumed to be a very basic condition for the creation of life since
it is the dust matter that constitutes galaxies, stars, planets, and
creatures including human beings.
%This can be viewed as a scientific
%alternative to the anthropic principle in some sense.
%Furthermore, due to
%the fact that the dust matter must be massive, our conjecture seems
%to indicate a new close tie between our large scale fiducial
%$\Lambda$CDM cosmological model and our small scale standard model
%or beyond for elementary particles, namely, the origin of mass for
%the baryonic matter and cold dark matter is determined to be
%intertwined with the present cosmological constant somehow or
%others.
\section{Conclusion}
Without knowledge of its microscopic makeup and specific dynamics,
the use of general principles to investigate a system can be very
rewarding. Examples of such principles include the widely accepted
thermodynamics and recently recognized holographic principle. The
holographic principle, albeit confirmed in limited concrete models,
is believed to be a law of physics that captures one of the most
crucial aspects of quantum gravity, and thus a key insight for guiding
the progressing search for a successful unified theory.

As a compelling pattern for the holographic principle, and an
equivalent formulation of generalized second law of thermodynamics
as well, our recently proposed covariant entropy conjecture has been
applied to cosmological models with the presence of a positive
cosmological constant. As a result, it is shown that our conjecture
places a severe constraint on acceptable cosmological models. Not
only does this conjecture rule out those cosmological models
disfavored by the anthropic principle, but also it imposes an upper
bound $10^{-60}$ on the cosmological constant for our own universe,
which thus opens an alternative macroscopic perspective to shed
light on the long-standing cosmological constant problem. Although
this upper bound does not fix the exceedingly tiny value of
cosmological constant, it is plausible that there are other
contributions to entropy density which we have not taken into
account, and a more complete picture of constituents and structures
of the universe, will yield a better estimate of the entropy density
and finally resolve the cosmological constant problem.
%In addition,
%as inferred by our conjecture, the presence of cosmological constant
%requires that during the evolution of the universe, elementary
%particles must acquire masses through Higgs mechanism or something
%deeper, to become constituents of dust matter, which obviously
%prefers a bio-friendly universe, thus provides a scientific
%alternative to the anthropic principle.

We conclude with an honest caveat. Although the results
obtained so far are particularly attractive as well as consistent
with our observational data, there remains a possibility that our
starting conjecture proves incorrect. It may be quite successful in
many respects only as a coincidence, but one should regard it as as
a warning, showing that our covariant entropy conjecture may require
additional justification and reformulation where it is violated,
rather than being a criterion for singling out acceptable models.
Therefore not only is it important to provide more indirect or peripheral justifications, 
but also it is needed to signify a deeper origin of our conjecture in an underlying quantum theory of gravity, 
such as causal set theory, loop quantum gravity, or string theory\cite{DS,RZ,Ashtekar,SV}.
\section*{Acknowledgements}
We are deeply indebted to Niayesh Afshordi, Achim Kempf, Federico
Piazza, Maxim Pospelov, Lee Smolin, Rafael Sorkin, and Leonard
Susskind for their helpful suggestions and comments. HZ would like
to express much gratitude to Stephen Hawking for his vivid
communication on the anthropic principle and related issues. He is
also grateful to Xiao Liu for valuable criticisms leading to improvement.
The work by SH was supported by NSFC(nos.10235040 and 10421003). HZ was
supported in part by the Government of China through
CSC(no.2007102530). This research was supported by Perimeter
Institute for Theoretical Physics. Research at Perimeter Institute
is supported by the Government of Canada through IC and by the
Province of Ontario through MRI.

\end{document}